\documentclass[runningheads]{llncs}
\usepackage{url}
\usepackage{amsmath}
\usepackage{amssymb}
\usepackage{enumitem} 
\usepackage{fancyhdr} 
\usepackage{listings}
\usepackage{lstcoq}
\usepackage{mathpartir}
\usepackage{hyperref}

\setlist[itemize]{topsep=1pt,itemsep=0ex,partopsep=1ex,parsep=1ex,itemindent=0pt,leftmargin=12pt}
\setlist[enumerate]{topsep=1pt,itemsep=0ex,partopsep=1ex,parsep=1ex,itemindent=0pt,leftmargin=12pt}

\newcommand{\evalto}{\mathrel{\Downarrow}}
\newcommand{\eval}[4]{\langle #1, #2 \rangle \evalto (#3, #4)}
\newcommand{\stepto}{\rightarrow}
\newcommand{\step}[2]{#1 \stepto #2}

\newcommand{\nil}{\texttt{[]}}
\newcommand{\cons}{\mathbin{::}}
\newcommand{\app}{\mathbin{++}}
\newcommand{\Perm}{\text{Perm}}

\begin{document}

\title{Smart Contract Interactions in Coq}

\author{Jakob Botsch Nielsen\orcidID{0000-0002-0459-2678} \and Bas
Spitters\orcidID{0000-0002-2802-0973}}
\institute{Concordium Blockchain Research Center, Computer Science, Aarhus University
\email{\{botsch,spitters\}@cs.au.dk}}

\maketitle

\begin{abstract}
We present a model/executable specification of smart contract execution in Coq.
Our formalization allows for inter-contract communication and generalizes
existing work by allowing modelling of both depth-first execution blockchains
(like Ethereum) and breadth-first execution blockchains (like Tezos). We
represent smart contracts programs in Coq's functional language Gallina, enabling easier reasoning about
functional correctness of concrete contracts than other approaches. In
particular we develop a Congress contract in this style. This contract -- a
simplified version of the infamous DAO -- is interesting because of its very
dynamic communication pattern with other contracts. We give a high-level partial
specification of the Congress's behavior, related to reentrancy, and prove that
the Congress satisfies it for all possible smart contract execution orders.

\end{abstract}

\section{Introduction}
Since Ethereum, blockchains make a clear separation between the consensus layer
and the execution of interacting smart contracts. In Ethereum's Solidity
language contracts can arbitrarily call into other contracts as regular function
calls. Modern blockchains further separate the top layer in an execution layer
and a contract layer. The execution layer schedules the calls between the
contracts and the contract layer executes individual programs. The choice of
execution order differs between blockchains. For example, in Ethereum the
execution is done in a synchronous (or depth first) order: a call completes
fully before the parent continues, and the parent is able to observe its result.
Tezos and Scilla use a breadth first order instead, where observing the result
is not possible.

We
provide\footnote{\url{https://gitlab.au.dk/concordium/smart-contract-interactions/tree/v1.0}}
a model/executable specification of the execution and contract layer of a third
generation blockchain in the Coq proof assistant. We use Coq's embedded
functional language Gallina to model contracts and the execution layer. This
language can be extracted to certified programs in for example Haskell or Ocaml.
Coq's expressive logic also allows us to write concise proofs. The consensus
protocol provides a consistent global state which we treat abstractly in our
formalization.

We work with an account-based model. We could also model the UTxO model by
converting a list of UTxO transactions to a list of account
transactions~\cite{account_vs_utxo}. Like that work, we do not model the
cryptographic aspects, only the accounting aspects: the transactions and
contract calls.

Section~\ref{section:implementation} describes the implementation of
the execution layer in Coq. In Section~\ref{section:congress} we provide a simple principled specification for the
Congress. By using such specifications one avoids having to deal with
reentrancy bugs in a post-hoc way. Section~\ref{related}
discusses related and future work. 


\section{Implementation}\label{section:implementation}

\subsection{Basic assumptions}
Our goal is to model execution of smart contracts. To do so we will require some
basic operations that are to be used both by smart contracts and when specifying
our semantics. We do this with a typeclass in Coq:

\begin{coq-small}
Class ChainBase :=
  { Address : Type;
    address_countable :> Countable Address;
    address_is_contract : Address -> bool; ... }.
\end{coq-small}
We require a countable \coqe{Address} type with a clear separation between
addresses belonging to contracts and to users. While this separation is not
provided in Ethereum its omission has led to exploits before\footnote{See for
instance
\url{https://www.reddit.com/r/ethereum/comments/916xni/how_to_pwn_fomo3d_a_beginners_guide/}}
and we view it as realistic that future blockchains allow this. Other
blockchains commonly provide this by using some specific format for contract
addresses, for example, in Bitcoin such pay-to-script-hash addresses always
start with $3$.

All semantics and smart contracts will be abstracted over an instance of this
type, so in the following sections we will assume we are given such an instance.

\subsection{Smart Contracts}
We will consider a pure functional smart contract language. Instead of
modelling the language as an abstract syntax tree in Coq, as
in~\cite{deep_shallow_embeddings}, we model individual smart contracts
as records with (Coq) functions.

\subsubsection{Local state.} It is not immediately clear how to represent smart
contracts by functions. For one, smart contracts have local state that they
should be able to access and update during execution. In Solidity, the language
typically used in Ethereum, this state is mutable and can be changed at any
point in time. It is possible to accomplish something similar in pure languages,
for example by using a state monad, but we do not take this approach. Instead we
use a more traditional approach where the contract takes as input its state and
returns an updated state which is similar to Liquidity.

Different contracts will typically have different types of states. A
crowdfunding contract may wish to store a map of backers in its state while an
auction contract would store information about ongoing auctions. To facilitate
this polymorphism we use an intermediate storage type called
\coqe{SerializedValue}. We define conversions between \coqe{SerializedValue} and
primitive types like booleans and integers plus derived types like pairs, sums
and lists. Additionally we provide Coq tactics that can automatically generate
conversions for custom user types like inductives and records. This allows
conversions to be handled implicitly and mostly transparently to the user.

\subsubsection{Inter-contract communication.} In addition to local state we also
need some way to handle inter-contract communication. In Solidity contracts can
arbitrarily call into other contracts as regular function calls. This would once
again be possible with a monadic style, for example by the use of a promise
monad where the contract would ask to be resumed after a call to another
contract had finished. To ease reasoning we choose a simpler approach where
contracts return actions that indicate how they would like to interact with the
blockchain, allowing transfers, contract calls and contract deployments only at
the end of (single steps of) execution. The blockchain will then be responsible
for scheduling these actions in what we call its execution layer.

With this design we get a clear separation between contracts and their
interaction with the chain. That such separations are important has been
realized before, for instance in the design of Michelson and
Scilla~\cite{scilla}. Indeed, a "tail-call" approach like this forces the
programmer to update the contract's internal state before making calls to other
contracts, mitigating by construction reentrancy issues such as the infamous DAO
exploit.

Thus, contracts will take their local state and some data allowing them to query
the blockchain. As a result they then optionally return the new state and some
operations (such as calls to other contract) allowing inter-contract
communication. Tezos' Michelson language follows a similar approach.

The Ethereum model may be compared to object-oriented programming. Our model is
similar to the actor model as contracts do not read or write the state of
another contract directly, but instead communicate via messages instead
of shared memory. Liquidity and the IO-automata-based Scilla use similar models.

\subsubsection{The contract.} Smart contracts are allowed to query various data
about the blockchain. We model this with a data type:

\begin{coq-small}
Definition Amount := Z.
Record Chain := { chain_height : nat;
                  current_slot : nat;
                  finalized_height : nat;
                  account_balance : Address -> Amount; }.
\end{coq-small}

We allow contracts to access basic details about the blockchain, like the
current chain height, slot number and the finalized height. The slot number is
meant to be used to track the progression of time; in each slot, a block can be
created, but it does not have to be. The finalized height allows contracts to
track the current status of the finalization layer available in for example the
Concordium blockchain~\cite{afjort}. This height is different from the chain
height in that it guarantees that blocks before it can not be changed. We
finally also allow the contract to access balances of accounts as is common
in other blockchains. In sum, the following data types model the contracts:

\begin{coq-small}
Record ContractCallContext :=
  { ctx_from : Address;
    ctx_contract_address : Address;
    ctx_amount : Amount; }.
Inductive ActionBody :=
| act_transfer (to : Address) (amt : Amount)
| act_call (to : Address) (amt : Amount) (msg : SerializedValue)
| act_deploy (amt : Amount) (c : WeakContract) (setup : SerializedValue)
with WeakContract :=
| build_weak_contract
  (init : Chain -> ContractCallContext -> SerializedValue (* setup *)
          -> option SerializedValue)
  (receive : Chain -> ContractCallContext -> SerializedValue (* state *)
             -> option SerializedValue (* message *)
             -> option (SerializedValue * list ActionBody)).
\end{coq-small}

Here the \coqe{ContractCallContext} provides the contract with information about
the transaction that resulted in its execution. It contains the source address
(\coqe{ctx_from}), the contract's own address (\coqe{ctx_contract_address}) and
the amount of money transferred (\coqe{ctx_amount}). The \coqe{ActionBody} type
represents operations that interact with the chain. It allows for
messageless transfers (\coqe{act_transfer}), calls with messages
(\coqe{act_call}), and deployment of new contracts (\coqe{act_deploy}). The
contract itself is two functions. The \coqe{init} function is used when a
contract is deployed to set up its initial state, while the \coqe{receive}
function will be used for transfers and calls with messages afterwards. They
both return option types, allowing the contract to signal invalid calls or
deployments. The \coqe{receive} function additionally returns a list of
\coqe{ActionBody} that it wants to be scheduled, as we described earlier. This
data type does not contain a source address since it is implicitly the
contract's own address. Later, we will also use a representation where there
\textit{is} a source address; we call this type \coqe{Action}:

\begin{coq-small}
Record Action := { act_from : Address; act_body : ActionBody; }.
\end{coq-small}
This type resembles what is normally called a transaction, but we make a
distinction between the two. An \coqe{Action} is a \textit{request} by a
contract or external user to perform some operation. When executed by an
implementation, this action will affect the state of the blockchain in some way.
It differs from transactions since \coqe{act_deploy} does not contain the
address of the contract to be deployed. This models that it is the
implementation that picks the address of a newly deployed contract, not the
contract making the deployment. We will later describe our
\coqe{ActionEvaluation} type which captures more in depth the choices made by
the implementation while executing an action.

The functions of contracts are typed using the \coqe{SerializedValue} type. This
is also the reason for the name \coqe{WeakContract}. This makes specifying
semantics simpler, since the semantics can deal with contracts in a generic way
(rather than contracts abstracted over types). However, this form of
"string-typing" makes things harder when reasoning about contracts. For this
reason we provide a dual notion of a \textit{strong} contract, which is a
polymorphic version of contracts generalized over the setup, state and message
types. Users of the framework only need to be aware of this notion of contract,
which does not contain references to \coqe{SerializedValue} at all.

One could also imagine an alternative representation using a dependent record of
setup, state and message types plus functions over those types. However, in such
a representation it is unclear how to allow contracts to send messages to other
contracts when the blockchain itself does not have any knowledge about concrete
contracts.

\subsection{Semantics of the execution layer}

\subsubsection{Environments.} The \coqe{Chain} type shown above is merely the
contract's view of the blockchain and does not store enough information to allow
the blockchain to run actions. More specifically we need to be able to look up
information about currently deployed contracts like their functions and state.
We augment the \coqe{Chain} type with this information and call it an
\coqe{Environment}:

\begin{coq-small}
Record Environment :=
  { env_chain :> Chain;
    env_contracts : Address -> option WeakContract;
    env_contract_states : Address -> option SerializedValue; }.
\end{coq-small}

It is not hard to define functions that allow us to make updates to
environments. For instance, inserting a new contract is done by creating a new
function that checks if the address matches and otherwise uses the old map. In
other words we use simple linear maps in the semantics. In similar ways we can
update the rest of the fields of the \coqe{Environment} record.

\subsubsection{Evaluation of actions.} When contracts return actions the
execution layer will need to evaluate the effects of these actions. We define
this as a "proof-relevant" relation \coqe{ActionEvaluation} in Coq, with type
\coqe{Environment -> Action -> Environment -> list Action -> Type}. This
relation captures the requirements and effects of executing the action in the
environment. It is "proof-relevant", meaning that the choices made by the
execution layer can be inspected. For example, when an action requests to deploy
a new contract, the address selected by the implementation can be extracted from
this relation.

We define the relation by three cases: one for transfers of money, one for
deployment of new contracts, and one for calls to existing contracts. To
exemplify this relation we give its formal details for the simple transfer case
below:

\begin{coq-small}
| eval_transfer :
    forall {pre : Environment} {act : Action} {new_env : Environment}
           (from to : Address) (amount : Amount),
      amount <= account_balance pre from ->
      address_is_contract to = false ->
      act_from act = from ->
      act_body act = act_transfer to amount ->
      EnvironmentEquiv new_env (transfer_balance from to amount pre) ->
      ActionEvaluation pre act new_env []
\end{coq-small}

In this case the sender must have enough money and the recipient cannot be a
contract. When this is the case a transfer action and the old environment
evaluate to the new environment where the \coqe{account_balance} has been
updated appropriately. Finally, such a transfer does not result in more actions
to execute since it is not associated with execution of contracts. Note that we
close the evaluation relation under extensional equality
(\coqe{EnvironmentEquiv}).

We denote this relation by the notation $\eval{\sigma}{a}{\sigma'}{l}$. The
intuitive understanding of this notation is that evaluating the action $a$ in
environment $\sigma$ results in a new environment $\sigma'$ and new actions to
execute $l$.

\subsubsection{Chain traces.} The \coqe{Environment} type captures enough
information to evaluate actions. We further augment this type to keep track of
the queue of actions to execute. In languages like Solidity this data is encoded
implicitly in the call stack, but since interactions with the blockchain are
explicit in our framework we keep track of it explicitly.

\begin{coq-small}
Record ChainState := { chain_state_env :> Environment;
                       chain_state_queue : list Action; }.
\end{coq-small}

We now define what it means for the chain to take a step.
Formally, this is defined as a "proof-relevant" relation \coqe{ChainStep} of
type \coqe{ChainState -> ChainState -> Type}. We denote this relation with the
notation $\step{(\sigma, l)}{(\sigma', l')}$, meaning that we can step from the
environment $\sigma$ and list of actions $l$ to the environment $\sigma'$ and
list of actions $l'$. We give this relation as simplified inference rules:

\begin{figure}\vspace{-2.3em}
\begin{mathpar}
  \infer[step-block]
  { b \ \text{valid for} \ \sigma \\ acts \ \text{from users} }
  { \step{(\sigma, \nil)}{(\texttt{add\_block}\ b\ \sigma, acts)} }
  \and
  \infer[step-action]
  { \eval{\sigma}{a}{\sigma'}{l} }
  { \step{(\sigma, a \cons l')}{(\sigma', l \app l')} }
  \and
  \infer[step-permute]
  { \Perm(l, l') }
  { \step{(\sigma, l)}{(\sigma, l')} }
\end{mathpar}
\vspace{-3.0em}\end{figure}

The \textsc{step-block} rule allows the addition of a new block ($b$) containing
some actions ($acts$) to execute. We require that the block is valid for the
current environment (the "$b$ valid for $\sigma$" premise), meaning that it needs
to satisfy some well-formedness conditions. For example, if the chain currently
has height $n$, the next block added needs to have height $n + 1$. There are
other well-formedness conditions on other fields, such as the block creator, but
we omit them here for brevity. Another condition is that all added actions must
come from users (the "$acts$ from users" premise). This models the real world
where transactions added in blocks are "root transactions" from users. This
condition is crucial to ensure that transfers from contracts can happen only due
to execution of their associated code. When the premises are met we update
information about the current block (such as the current height and the balance
of the creator, signified by the \texttt{add\_block} function) and update that
the queue now contains the actions that were added.

The \textsc{step-action} rule allows the evaluation of the first action in the
queue, replacing it with the resulting new actions to execute. This new list
($l$ in the rule) is concatenated at the beginning, corresponding to using the
queue as a stack. This results in a depth-first execution order of actions. The
\textsc{step-permute} rule allows an implementation to use a different order of
reduction by permuting the queue at any time. For example, it is possible to
obtain a breadth-first order of execution by permuting the queue so that newly
added events are in the back. In this case the queue will be used like an actual
FIFO queue.

Building upon steps we can further define \textit{traces} as the proof-relevant
reflexive transitive closure of the step relation. In other words, this is a
sequence of steps where each step starts in the state that the previous step
ended in. Intuitively the existence of a trace between two states means that
there is a semantically correct way to go between those states. If we let
$\varepsilon$ denote the empty environment and queue this allows us to define a
concept of \textit{reachability}. Formally we say a state $(\sigma, l)$ is
\textit{reachable} if there exists a trace starting in $\varepsilon$ and ending
in $(\sigma, l)$.
%
%
Generally, only reachable states are interesting to consider and most proofs are
by induction over the trace to a reachable state.

\subsection{Building blockchains}
We connect our semantics to an executable definition of a blockchain with a
typeclass in Coq:
\begin{coq-small}
Class ChainBuilderType := {
 builder_type : Type;
 builder_initial : builder_type;
 builder_env : builder_type -> Environment;
 builder_add_block (builder : builder_type) (header : BlockHeader)
  (actions : list Action) : option builder_type;
 builder_trace (builder : builder_type) :
  ChainTrace empty_state (build_chain_state (builder_env builder) []); }.
\end{coq-small}
A chain builder is a dependent record consisting of an implementation type
(\coqe{builder_type}) and several fields using this type. It must provide an
initial builder, which typically would be an empty chain, or a chain containing
just a genesis block. It must be convertible to an environment allowing to query
information about the state. It must define a function that allows addition of
new blocks. Finally, the implementation needs to be able to give a trace showing
that the current environment is reachable with no more actions left in the queue
to execute. This trace captures a definition of soundness, since it means that
the state of such a chain builder will always be reachable.

\subsubsection{Instantiations.} We have implemented two instances of the
\coqe{ChainBuilderType} typeclass. Both of these are based on finite maps from
the std++ library used by Iris~\cite{iris} and are thus relatively efficient
compared to the linear maps used to specify the semantics. The difference in the
implementations lies in their execution model: one implementation uses a
depth-first execution order, while the other uses a breadth-first execution
order. The former execution model is similar to the EVM while the latter is
similar to Tezos and Scilla.

These implementations are useful as sanity checks but they also serve other
useful purposes in the framework. Since they are executable they can be used to
test concrete contracts that have been written in Coq. This involves writing the
contracts and executing them using Coq's \coqe{Compute} vernacular to inspect
the results. In addition, they can also be used to give counter-examples to
properties. In the next section we will introduce the \textit{Congress}
contract, and we have used the depth-first implementation of our semantics to
formally show that this contract with a small change can be exploited with
reentrancy.

\section{Case: Congress -- a simplified DAO}\label{section:congress}
In this section we will present a case study of implementing and partially
specifying a complex contract in our framework.

\subsubsection{The Congress contract}
Wang~\cite{timl_phd_thesis} gives a list of eight interesting Ethereum
contracts. One of these is the \textit{Congress} in which
members of the contract vote on \textit{proposals}. Proposals contain
transactions that, if the proposal succeeds, are sent out by the Congress. These
transactions are typically monetary amounts sent out to some address, but they
can also be arbitrary calls to any other contract.

We pick the Congress contract because of its complex dynamic interaction
with the blockchain and because of its similarity to the infamous DAO contract
that was deployed on the Ethereum blockchain and which was eventually hacked by
a clever attacker exploiting reentrancy in the EVM. The Congress can be seen as
the core of the DAO contract, namely the proposal and voting mechanisms.

We implement the logic of the Congress in roughly 150 lines of Coq code.
The type of messages accepted by the Congress can be thought of as its interface
since this is how the contract can be interacted with:

\begin{coq-small}
Inductive Msg :=
  | transfer_ownership : Address -> Msg
  | change_rules : Rules -> Msg
  | add_member : Address -> Msg
  | remove_member : Address -> Msg
  | create_proposal : list CongressAction -> Msg
  | vote_for_proposal : ProposalId -> Msg
  | vote_against_proposal : ProposalId -> Msg
  | retract_vote : ProposalId -> Msg
  | finish_proposal : ProposalId -> Msg.
\end{coq-small}

The Congress has an owner who is responsible for managing the rules of the
Congress and the member list. By default, we set this to be the creator of the
Congress. The owner can transfer his ownership away with the
\coqe{transfer_ownership} message. It is possible to make the Congress its own
owner, in which case all rule changes and modifications to the member list must
happen through proposals (essentially making the Congress a democracy).

Anyone can use the \coqe{create_proposal} and \coqe{finish_proposal} messages.
We allow proposals to contain any number of actions to send out, though we
restrict the actions to only transfers and contract calls (i.e. no contract
deployments). This restriction is necessary because deployments would require
the state of the Congress to contain the contracts to deploy. Since contracts
are functions in our shallow embedding this would require storing higher order
state which we do not allow in the framework. This is a downside to the shallow
embedding -- with a deep embedding like \cite{deep_shallow_embeddings}, the code
could be stored as an AST or bytes.

The rules of the Congress specify how long proposals need to be debated. During
this period, members of the Congress have the ability to vote on the proposal.
Once debated, a proposal can be finished and the Congress will remove it from
its internal storage and send out its actions if it passed.

\subsubsection{A partial specification}
The DAO vulnerability was in reward payout code in which a specially crafted
contract could reenter the DAO causing it to perform actions an unintended
number of times. Specifically, the attacker was able to propose a so-called
\textit{split} and have the original DAO transfer a disproportionate amount of
money to a new DAO contract under his control. Congress does not allow splits,
but the same kind of bug would be possible in code responsible for carrying out
proposals.

Previous research such as \cite{evm-fstar} has focused on defining this kind of
reentrancy formally. Such (hyper-)properties are interesting, but they also rely
heavily on the benefit of hindsight and their statements are complex and hard to
understand. Instead we would like to come up with a natural specification for
the Congress that a programmer could reasonably have come up with, even without
knowledge of reentrancy or the exploit. Our goal with this is to apply the
framework in a very concrete setting.

The specification we give is based on the following observation: any transaction
sent out by the Congress should correspond to an action that was previously
created with a \coqe{create_proposal} message. This is a temporal property
because it says something about the past whenever an outgoing transaction is
observed. Temporal logic is not natively supported by Coq, so this would require
some work. Therefore we prefer a similar but simpler property: the number of
actions in previous \coqe{create_proposal} messages is always greater than or
equal to the total number of transactions the Congress has sent out. Our main result about the Congress is a formal proof
that this always holds after adding a block:

\begin{coq-small}
Corollary congress_txs_after_block {ChainBuilder : ChainBuilderType}
          prev creator header acts new :
  builder_add_block prev creator header acts = Some new ->
  forall addr,
   env_contracts new addr = Some (Congress.contract : WeakContract) ->
   length (outgoing_txs (builder_trace new) addr) <=
   num_acts_created_in_proposals (incoming_txs (builder_trace new) addr).
\end{coq-small}

This result states that, after adding a block, any address at which a Congress
contract is deployed satisfies the property previously described. The number of
actions created in previous \coqe{create_proposal} messages is calculated by
function \coqe{num_acts_created_in_proposals}. The \coqe{incoming_txs} and
\coqe{outgoing_txs} functions are general functions that finds transactions
(evaluation of actions) in a trace. In this sense the property treats the
contract as a black box, stating only things about the transactions that have
been observed on the blockchain.

This is not a full specification of the behavior of the Congress but proving
this property can help increase trust in the Congress. In particular it would
not have been provable in the original DAO contract because of the reentrancy
exploit where the DAO sent out an unbounded number of transactions. Note
also that we do not want to exclude reentrancy entirely: indeed, in the
situation where the Congress is its own owner reentrancy is required for
changing rules and the member list.

We prove the property by generalizing it over the following data:

\begin{itemize}
  \item The internal state of the contract; more specifically, the current
    number of actions in proposals stored in the internal state.
  \item The number of transactions sent out by the Congress, as before.
  \item The number of actions \textit{in the queue} where the Congress
    is the source.
  \item The number of actions created in proposals, as before.
  \end{itemize}
This results in a stronger statement from which the original result follows. The
key observations are that
\begin{enumerate}
  \item\label{item:create_proposal} When a proposal is created, the number of
    actions created in proposals goes up, but so does the number of actions in
    the internal state of the Congress.

  \item\label{item:execute_proposal} When a proposal is finished, the number of
    actions in the internal state goes down, but the number of actions in the
    queue goes up accordingly (assuming the proposal was voted for). In other
    words, actions "move" from the Congress's internal state to the queue.

  \item\label{item:temporal} When an outgoing transaction appears on the chain
    it is because an action moved out of the queue.
\end{enumerate}
Especially observation~\ref{item:temporal} is interesting. It allows us to
connect the evaluation of a contract in the past to its resulting transactions
on the chain, even though these steps can be separated by many unrelated steps
in the trace.

The proof of the stronger statement is straightforward by inducting over the
trace. When deploying the Congress we need to establish the invariant which
boils down to proving functional correctness of the \coqe{init} function and the
use of some results that hold for contracts which have just been deployed (for
instance, such contracts have not made any outgoing transactions). On calls to
the Congress the invariant needs to be reestablished, which boils down to
proving functional correctness of the \coqe{receive} function. Crucially, we can
reestablish the invariant because the implementation of the Congress clears out
proposals from its state \textit{before} the actions in the proposal are
evaluated (the DAO was vulnerable because it neglected to do this on splits).

\section{Conclusions}\label{related}\label{conclusion}
We have formalized the execution model of blockchains in Coq and used our
formalization to prove formally a result about a concrete contract. Our
formalization of blockchain semantics is flexible in that it accounts both for
depth-first and breadth-first execution order, generalizing existing blockchains
and previous work, while remaining expressive enough to allow us to prove
results about complex contracts. We showed for a Congress -- a simplified
version of the DAO, which still has a complex dynamic interaction pattern --
that it will never send out more transactions than have been created in
proposals. This is a natural property that aids in increasing trust that this
contract is not vulnerable to reentrancy like the DAO.

\subsubsection{Related work}Both Simplicity~\cite{Simplicity} and Scilla~\cite{scilla} are smart contract
languages with an embedding in Coq. Temporal properties of several
smart contracts have been verified in Scilla~\cite{Scilla-temporal}, although
our Congress contract is more complex than the contracts described in that
paper. We are unaware of an implementation of such a contract in Scilla. Scilla,
as an intermediate language which includes both a functional part and contract
calls, uses a CPS translation to ensure that every call to another contract is
done as the last instruction. In our model, the high-level language and the
execution layer are strictly separated.

The formalization of the EVM in F* \cite{evm-fstar} can be extracted and used to
run EVM tests to show that it is a faithful model of the EVM. However, they do
not prove properties of any concrete contracts. Instead they consider classes of
bugs in smart contracts and try to define general properties that prevent these.
One of these properties, call integrity, is motivated by the DAO and attempts to
capture reentrancy. Intuitively a contract satisfies call integrity if the calls
it makes cannot be affected by code of other contracts. VerX~\cite{VerX} uses
temporal logic and model checking to check a similar property. Such statements
are not hard to state in our framework given Coq's expressive logic, and it
seems this would be an appropriate property to verify for the Congress.
However, even a correct Congress does not satisfy this property, since it is
possible for called contracts to finish proposals which can cause the Congress
to perform calls. This property could potentially be proven in a version of the
Congress that only allowed proposals to be finished by humans, and not by
contracts.

\subsubsection{Future work}
More smart contracts are available in Wang's PhD thesis~\cite{timl_phd_thesis}
and specifying these to gain experience with using the framework will help
uncover how the framework itself should be improved. In this area it is also
interesting to consider more automatic methods to make proving more productive.
For example, temporal logics like LTL or CTL can be useful to specify properties
on traces and model checking these can be automated; see e.g.~\cite{VerX}.

Finally, while our current framework is inspired by and generalizes existing
blockchains, there is still more work to be done to get closer to practical
implementations. Gas is notoriously difficult to deal with in our shallow
embedding because tracking costs of operations can not be done automatically.
However, perhaps a monadic structure can be used
here~\cite{monadic_complexity_coq}. We have connected our work with a deep
embedding of a functional language~\cite{annenkov2019smart} and explored pros
and cons of shallow and deep embeddings in that work. We plan to use this deep
embedding to explore reasoning about gas. In the other direction it is
interesting to consider extraction of the execution layers we have shown to
satisfy our semantics and extraction of verified contracts into other languages
like Liquidity, Oak or Solidity.

\subsubsection{Acknowledgements.} We would like to thank the Oak team for
discussions.

\bibliography{literature}

\begin{thebibliography}{10}
\providecommand{\url}[1]{\texttt{#1}}
\providecommand{\urlprefix}{URL }
\providecommand{\doi}[1]{https://doi.org/#1}

\bibitem{annenkov2019smart}
Annenkov, D., Nielsen, J.B., Spitters, B.: Towards a smart contract
  verification framework in {Coq}. arXiv:1907.10674  (2019)

\bibitem{deep_shallow_embeddings}
Annenkov, D., Spitters, B.: Deep and shallow embeddings in {Coq}. TYPES  (2019)

\bibitem{evm-fstar}
Grishchenko, I., Maffei, M., Schneidewind, C.: A semantic framework for the
  security analysis of ethereum smart contracts. In: PoST. pp. 243--269.
  Springer (2018)

\bibitem{iris}
Jung, R., Krebbers, R., Jourdan, J.H., Bizjak, A., Birkedal, L., Dreyer, D.:
  Iris from the ground up: A modular foundation for higher-order concurrent
  separation logic. Journal of Functional Programming  \textbf{28} (2018)

\bibitem{afjort}
Magri, B., Matt, C., Nielsen, J.B., Tschudi, D.: Afgjort -- a semi-synchronous
  finality layer for blockchains.
  \href{https://eprint.iacr.org/2019/504}{Cryptology ePrint 2019/504} (2019)

\bibitem{monadic_complexity_coq}
McCarthy, J., Fetscher, B., New, M.S., Feltey, D., Findler, R.B.: A {Coq}
  library for internal verification of running-times. Science of Computer
  Programming  \textbf{164},  49--65 (2018)

\bibitem{Simplicity}
O'Connor, R.: Simplicity: {A} new language for blockchains. arXiv:1711.03028
  (2017)

\bibitem{VerX}
Permenev, A., Dimitrov, D., Tsankov, P., Drachsler-Cohen, D., Vechev, M.: Verx:
  Safety verification of smart contracts. Security and Privacy 2020  (2019)

\bibitem{scilla}
Sergey, I., Kumar, A., Hobor, A.: Scilla: a smart contract intermediate-level
  language. arXiv:1801.00687  (2018)

\bibitem{Scilla-temporal}
Sergey, I., Kumar, A., Hobor, A.: Temporal properties of smart contracts. In:
  Margaria, T., Steffen, B. (eds.) Leveraging Applications of Formal Methods,
  Verification and Validation. Industrial Practice. pp. 323--338. Springer
  (2018)

\bibitem{timl_phd_thesis}
Wang, P.: Type System for Resource Bounds with Type-Preserving Compilation.
  Ph.D. thesis, MIT (2018)

\bibitem{account_vs_utxo}
Zahnentferner, J.: Chimeric ledgers: Translating and unifying {UTXO}-based and
  account-based cryptocurrencies.
  \href{https://eprint.iacr.org/2018/262}{Cryptology ePrint 2018/262} (2018)

\end{thebibliography}

\end{document}